\documentclass[aps,final,notitlepage,oneside,twocolumn,nobibnotes,nofootinbib,
superscriptaddress,noshowpacs,centertags]{revtex4-1}

\usepackage[utf8]{inputenc}
\usepackage[russian]{babel}
\usepackage{graphicx}
\usepackage{latexsym}
\usepackage{amssymb}
\usepackage{amsmath}
\usepackage{float}
\usepackage{color}

\begin{document}
	
\title{
Absorption in the 21 cm Hydrogen Line at $z>10$ as a Sensitive Tool\\
for the Construction of a Cosmological Model on Small Scales}
	\author{Yu.N. Eroshenko}\thanks{e-mail: eroshenko@inr.ac.ru}
	\affiliation{Institute for Nuclear Research, Russian Academy of Sciences, Moscow, 117312 Russia}
	\author{V.N. Lukash}\thanks{e-mail: lukash@asc.rssi.ru}
	\affiliation{Astro Space Center, Lebedev Physical Institute, Russian Academy of Sciences, Moscow, 117997 Russia}
	\author{E.V. Mikheeva}\thanks{e-mail: helen@asc.rssi.ru}
	\affiliation{Astro Space Center, Lebedev Physical Institute, Russian Academy of Sciences, Moscow, 117997 Russia}
	\author{S.V. Pilipenko}\thanks{e-mail: spilipenko@asc.rssi.ru}
	\affiliation{Astro Space Center, Lebedev Physical Institute, Russian Academy of Sciences, Moscow, 117997 Russia}
	\author{M.V. Tkachev}\thanks{e-mail: mtkachev@asc.rssi.ru}
	\affiliation{Astro Space Center, Lebedev Physical Institute, Russian Academy of Sciences, Moscow, 117997 Russia}
	
\begin{abstract}
The cosmic microwave background absorption intensity in the 21~cm line of neutral hydrogen in the presence of additional power in the form of a “bump” in the spectrum of cosmological density perturbations is calculated. The main absorption-amplifying effect is the earlier birth of the first stars forming an ultraviolet radiation background. This radiation reduces the spin temperature of neutral hydrogen and, thus, amplifies the absorption in the 21 cm line. A comparison of various cosmological models (with and without a bump in the density perturbation spectrum) shows that it is possible to determine the probable position of the bump in the perturbation spectrum and, thus, to reconstruct the spectrum of cosmological perturbations on scales $k>1$~Mpc${}^{-1}$ from the position of the absorption frequency profile. 
\vskip 5mm

\\
\noindent
Astronomy Letters, 2025, Vol. 51, No. 4, pp. 189–197; DOI: 10.1134/S1063773725700343
\end{abstract}
	
	\maketitle 
			
	\section{INTRODUCTION}
The 21~cm line of neutral hydrogen arises during the transitions in the hydrogen atom between the hyperfine splitting states with a total atomic spin of 0 and 1 (for a review, see Pritchard and Loeb 2012). Observations of this line provide information about a number of important astrophysical and cosmological processes at redshifts z > 6 that affected its formation, including the formation of the first stars and galaxies as well as black holes that are born, for example, during the direct collapse of gas clouds (Zhang et al. 2025). Investigating the absorption and emission in the 21~cm line of neutral hydrogen can become an extremely useful tool for reconstructing the spectrum of cosmological density perturbations at $k>1$~Mpc$^{-1}$ after this line will be reliably detected with radio telescopes when observing the Universe at $z>10$.

The cosmological epoch of Dark Ages lasted from hydrogen recombination to reionization in the Universe, i.e., at redshifts $z\sim6.4-1100$. The “spin” temperature important for the 21~cm transitions, which determines the relative populations of the upper and lower levels of hydrogen atoms, at $z>30$ was determined by the collisions of atoms with one another and with electrons. At lower z the collisions became inefficient and the spin temperature tended to the cosmic microwave background (CMB) temperature, which was higher than the kinetic temperature of the baryonic gas due to the difference in the cooling laws of the gases of relativistic and nonrelativistic particles as the Universe expanded. As a result, at $z\leq30$, but before the formation of the first stars, there was no absorption in the 21~cm line. Finally,
when stars began to appear within dark matter halos, their ultraviolet (UV) radiation reduced the spin temperature due to the Wouthuysen–Field (WF) effect (Wouthuysen 1952; Field 1958), establishing a coupling between the spin and kinetic degrees of freedom. However, the spin temperature probably had no time to reach the kinetic temperature of the baryonic gas, while the gas itself was slightly heated and ionized by the radiation of early stars and shocks. Nevertheless, the reduction in spin temperature turned out to be sufficient for the appearance of a deep absorption trough near $z\sim17-25$ (the exact value depends on the spectrum of cosmological density perturbations). After the gas heating and ionization, the absorption was replaced by the emission in this line.

Two experiments to observe the absorption line have been performed to date. The observed wavelength is shifted to $21\times(1+z)$~cm through the cosmological expansion. Observations with the EDGES radio telescopes (Monsalve et al. 2017, 2018, 2019; Bowman et al. 2018) revealed an “absorption trough” whose position at $z\sim17$ generally corresponds to the theoretical expectations within the standard $\Lambda$CDM model, but the absorption depth is several times larger than follows from the calculations of Bowman et al. (2018). If the EDGES observational data are valid, then it is necessary to invoke new physical effects to explain them, since the included processes are probably unable to explain the so strong absorption. For example, a model for the interaction of baryons with dark matter particles, whereby the baryonic gas cools down approximately by half, was proposed (Barkana 2018). This could reduce the spin temperature and, accordingly, amplify the absorption. Note, however, that the second SARAS~3 experiment (Singh et al. 2022; Bevins et al. 2022) failed to detect the absorption trough. Therefore, both the real depth of the absorption trough and its very presence so far remain in question.

The main factor affecting the spin temperature is the formation of the first stars. The first stars are formed within fairly massive dark matter halos, in which the virial temperature is sufficient to launch the processes of gas cooling and its compression followed by fragmentation. The progress of the formation of dark matter halos depends significantly on the shape of the cosmological perturbation spectrum on the corresponding scales. The goal of this paper is to investigate the influence of the density perturbation spectrum on the absorption in the 21~cm line, and, in future, with the availability of reliable observational data on the absorption, the inverse problem can be considered—obtaining information about the power spectrum from the observations of the absorption trough shape. We assume that the spectrum of cosmological density perturbations has a standard form, but with an additional narrow peak (bump) on some scale. The influence of such a bump on the formation of early galaxies was considered in Padmanabhan and Loeb (2023) and Tkachev et al. (2024a), where it was shown that more early galaxies are formed in this case, which can serve as an explanation of the James Webb Space Telescope (JWST) data.
	
Recently, Naik et al. (2025) computed the global absorption signal in the 21~cm line for a spectrum with an excess of power. The computation was performed using the \texttt{21cmFAST} package (Mesinger et al. 2011). The difference between the approaches used in this paper and in Naik et al. (2025) is as follows. Whereas in the \texttt{21cmFAST} package the linear theory of the growth of density perturbations with extrapolation in some cases to the weakly nonlinear region is used for the identification of virialized halos, we use the results of numerical N-body simulations for this purpose. Whereas the choice of the shape of the peak in the power spectrum was motivated by some inflation models, in our case the parameters of the peak (bump)
were chosen to explain the excess of galaxies in the early Universe observed by the JWST. The parameters of the bump in our case differ significantly from those used by Naik et al. (2025). We do not seek to accurately describe the shape of the absorption trough at low redshifts, since the processes occurring near the reionization epoch have not yet been completely clarified, as suggested, for example, by the ionizing photon overproduction crisis discussed in Munoz et al. (2024). We found a significant shift of the absorption trough boundary at high $z$ in the presence of a bump. Our calculations are qualitatively consistent with the conclusion of Naik et al. (2025) about the position of the absorption trough at high
redshifts.

A number of new radioastronomical observations in the meter band with various radio telescopes are expected in the near future. If these observations will allow one to adequately reveal the absorption trough and to describe its boundary, then this will give valuable information about the shape of the cosmological perturbation spectrum (in our model, about the position and height of the bump). In this case, the absorption in the 21~cm hydrogen line at $z>10$ will become an accurate tool for the construction of a cosmological model on small scales.

The formalism for calculating the global absorption in the 21~cm line was presented in Pritchard and Loeb (2012) and references therein. Below in the text of this paper we give the basic formulas to show the dependence of the results of our calculations on various parameters. Here, for our calculations we use the following cosmological parameters (Ade et al. 2014): $\Omega_{\Lambda}=0.69$, $\Omega_m=0.31$, $\Omega_b=0.048$, $h=0.67$, $n_s=0.96$.

	\section{INFLUENCE OF THE BUMP ON THE FORMATION OF THE FIRST STARS}
	\label{starsec}

When calculating the formation of large-scale structures, we assume that the standard $\Lambda$CDM perturbation spectrum is multiplied by an additional factor:   
        \begin{equation}
		1+A\times\exp\left(-\frac{(\log k-\log k_0)^2}{\sigma_0^2}\right),
		\label{bump}
	\end{equation}
where $A=20$, $k_0=4.69$~Mpc$^{-1}$, and $\sigma_0=0.1$ (following our earlier paper (Tkachev et al. 2024a), we designate this model as $gauss\_1$). The presence of a bump with such a shape leads to more efficient formation of galaxies in the early Universe and gives better agreement with the JWST observational data. The absence of a bump, i.e., the standard $\Lambda$CDM model, corresponds to $A=0$. We also investigated the cosmological model with a spectrum in which the bump had parameters $A=30$, $k_0=2.01$~Mпк$^{-1}$, and $\sigma_0=0.1$ (following the notation
introduced in Tkachev et al. (2024b), below we call this model $gauss\_k2\_A30$).

Note that the gravitational clustering process is affected not by the power spectrum itself, but by the root-mean-square (rms) density perturbation on a particular scale $R$. The variance of this quantity is defined by the expression
\begin{equation}
	\sigma_R^2 = \frac{1}{2\pi^2} \int_0^\infty k^2 P(k) W^2(kR) {\rm d}k\,,
	\label{sig0}
\end{equation}
where $W(kR)$ is the window function. The presence of a bump in the region of wave numbers $k_0$ leads to the fact that the variance (\ref{sig0}) is enhanced on scales $R\ll 1/k_0$ in a fairly universal way, since different Fourier components of a Gaussian random field are independent. Indeed, in this case we can approximately obtain
	\begin{equation}	\sigma_R\simeq\left[\tilde\sigma_R^2+B^2\right]^{1/2},
		\label{signew2}
	\end{equation}
	where 
	\begin{equation}
		B=\left(\frac{A\sigma_0 P(k_0)k_0^3}{2\pi^{3/2}}\right)^{1/2},
		\label{beq}
	\end{equation}
and $\tilde\sigma_R$ corresponds to $A=0$. The functional form (\ref{signew2}) is explained by the fact that the density perturbation is the sum of two independent variables with a Gaussian distribution. 

Mass scales $\sim10^8M_\odot$ corresponding to the condition $R\ll 1/k_0$ in the case under consideration are important for the effect of the formation of the first stars. The formation of stars will proceed in a similar way for bumps with different parameters if the values of (\ref{beq}) are equal for them. For example, almost equal $B$ are obtained for two bumps with parameters $A=20$, $k_0=4.69$~Mpc$^{-1}$, $\sigma_0=0.1$ and $A=30$, $k_0=2.01$~Mpc$^{-1}$, $\sigma_0=0.1$ and, therefore, the formation of stars and the absorption in the 21~cm line will be the same in these two cases. This conclusion is also qualitatively valid for a different bump shape if the bump region is localized on scales much larger than the scale of the halos in which cooling and star formation are possible. 

To investigate the evolution of dark matter, we performed three numerical N-body simulations in a cube with a volume of $(47\,\text{Mpc})^3$, each with $1024^3$ particles. One simulation corresponded to the $gauss\_1$ model and the second one corresponded to the standard $\Lambda CDM$ model; we also considered the $gauss\_k2\_A30$ model, where the bump was higher and was shifted to longer wavelengths. The size of the cube and the number of particles are the result of a compromise between a high resolution (the Nyquist frequency must exceed considerably the bump scale $k_0$, and the halos must contain at least several hundred particles) and a large size of the cube so that the fundamental perturbation mode (with the wavelength equal to the cube side) does not reach the nonlinear regime at $z=0$.

The initial conditions for our simulations were created at $z=120$ using the publicly accessible \texttt{ginnungagap} code\footnote{https://github.com/ginnungagapgroup/ginnungagap}; the matter power spectrum was determined for each simulation separately. It was generated using the publicly accessible \texttt{CLASS} code (Blas et al. 2011) for the standard $\Lambda$CDM model and using the function (\ref{bump}) for the models with a bump. The evolution of the density field was simulated with the \texttt{GADGET-2} code\footnote{http://wwwmpa.mpa-garching.mpg.de/~volker/gadget/} (Springel 2005), which is widely used to simulate the evolution of the structure of the Universe. The halos were analyzed using the \texttt{Rockstar} code\footnote{https://bitbucket.org/gfcstanford/rockstar} (Behroozi et al. 2013).

As was shown, for example, by Furlanetto (2006), Population II stars, which have already been enriched to some extent with metals, make a major contribution to the WF effect. Massive Population III stars preceding them, because of their high temperatures, have spectra shifted to high energies and make a smaller contribution. In this paper we will take into account only Population II stars. The (spatially) averaged luminosity of early short-lived stars is proportional to their formation rate. The latter, in turn, is proportional to the formation rate of virialized (collapsing) dark matter objects, i.e., the mean density of the stellar population
\begin{equation}
	\dot\rho_s=\bar\rho_b\times f_*\frac{df_{\rm coll}(z)}{dt}
\end{equation}
where the dot above the variable denotes a time derivative, $\bar\rho_b$ is the mean baryon density in the Universe, $f_*$ is the fraction of baryons in virialized halos converted into stars, and $f_{\rm coll}(z)$ is the fraction of collapsing mass (in virialized halos) as a function of the redshift. This quantity can be determined from the results of the previously described numerical simulations. We selected halos with virial temperatures $>10^4$~K in which efficient gas cooling is possible.

In the present-day Universe the mass fraction of baryons converted into stars is $f_*\sim0.1$. In many papers the same value is used to estimate the absorption in the 21~cm line. However, there are no strict arguments for such extrapolation. As the study of galaxies at $z\sim 4-8$ shows, $f_*$ can increase with increasing $z$ and can depend on the halo mass (Behroozi and Silk 2018). On the other hand, the problem of UV photon overproduction in early galaxies is well known (Munoz et al. 2024). In addition to the change in the escape fraction of UV photons, $f_{\rm esc}$, in the early Universe $f_*$ could also be different. In view of the existing great uncertainty, in this paper we will consider several $f_*$. Too large $f_*>0.1$ are unlikely to be possible, since this will aggravate the UV photon overproduction problem even more severely. Tkachev et al. (2024a) claim that in the presence of a bump smaller $f_*$ are required to explain the observed luminosity function of galaxies at high $z$ than those in the $\Lambda$CDM model.

\begin{figure}
	\centering
	\includegraphics[width=0.47\textwidth]{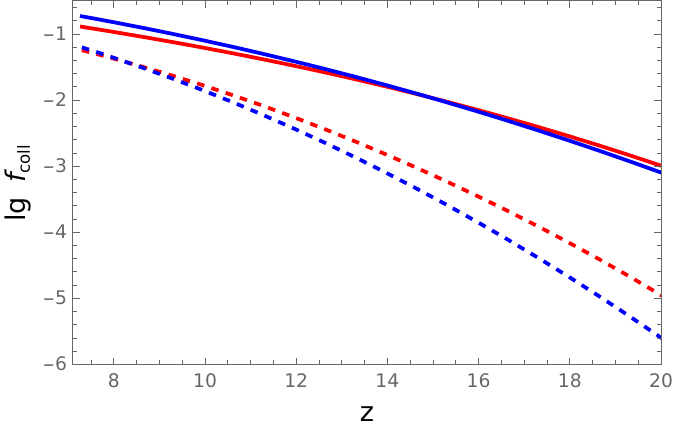}
	\caption{
Fraction of matter in the Universe in virialized dark matter halos versus redshift $z$. The red color indicates the curves constructed from the results of our numerical simulations; the criterion for the selection of halos was their virial temperature above $10^4$~K. The solid curve corresponds to the density perturbation spectrum with a bump; the dashed curve corresponds to the spectrum of the standard $\Lambda$CDM model. The blue color indicates the curves constructed from the Press–Schechter formalism for the minimum-mass boundary $M_b=10^8M_\odot$. The parameters of the bump are $A=20$, $k_0=4.69$~Mpc$^{-1}$ and  $\sigma_0=0.1$ (the $gauss\_1$ model).}
		\label{grcoll}
\end{figure}

Note that $f_{\rm coll}(z)$ found from our numerical simulations differs noticeably from the estimate based on the Press–Schechter theory (Press and Schechter 1974) with the minimum-mass boundary used, for example, by Furlanetto (2006):
\begin{equation}
f_{\rm coll}(z)={\rm erfc}\left\{\frac{\delta_c}{\sqrt{2}\sigma(M_b,z)}\right\},
\end{equation}
where $M_b\sim10^8M_\odot$ is the minimum mass of the dark matter halo in which stars are formed (according to the simplified criterion based on this mass boundary), and $\sigma(M_b,z)=\sigma(M_b)D(z)$ with the perturbation evolution factor $D(z)$ in the linear theory. The two approaches are compared in Fig.~\ref{grcoll}. It can be seen that the Press–Schechter theory for the models with a bump overestimates the number of halos by several times at low redshifts, but underestimates this number at high redshifts. At the same time, for two numerical models with different bumps, but having the same $B$, the fractions $f_\mathrm{coll}$ coincide, implying that this fraction is determined precisely by $B$ and not by the specific bump parameters.

	\section{EVOLUTION OF THE SPIN TEMPERATURE OF NEUTRAL HYDROGEN 
        AND THE GLOBAL ABSORPTION SIGNAL IN THE 21~cm LINE}
	\label{spinsec}

The global (sky-averaged) absorption in the 21~cm line in a uniformly distributed gas is calculated from the formula (Pritchard and Loeb 2012)
	\begin{equation}
		\delta T_b\simeq27(1-x)\left(1-\frac{T_\gamma(z)}{T_s(z)}\right)\left(\frac{1+z}{10}\right)\mbox{~мК},
		\label{dtfin}
	\end{equation}
where $x$ is the ionized hydrogen fraction, $T_\gamma$ is the CMB temperature, and $T_s(z)$ is the spin temperature dependent on the UV radiation of the first stars. We trace the behavior of the function $T_s(z)$ for two cases, in the presence and the absence of a bump, and calculate $\delta T_b$ for both cases.

The spin temperature is expressed as follows (for details, see Pritchard and Loeb 2012):    
	\begin{equation}
		T_s^{-1}=\frac{T_\gamma^{-1}+x_\alpha T_\alpha^{-1}+x_c T_K^{-1}}{1+x_\alpha+x_c},
		\label{ts1}
	\end{equation}
where $x_\alpha$ is the WF coupling coefficient due to the UV radiation of stars, $x_c$ is the coupling coefficient due to the thermal motion of atoms through collisions, and  $T_\alpha$ is the effective thermal radiation temperature that is equal to the kinetic gas temperature after the reradiation of photons with a good accuracy, $T_\alpha\approx T_K$. 

The coupling coefficient $x_\alpha$ is expressed as follows (Pritchard and Loeb 2012):
\begin{equation}
		x_\alpha=S_\alpha J_\alpha/J_\alpha^{C},
		\label{xaleq}
\end{equation} 
where $S_\alpha\sim1$ and $J_\alpha$ is the flux at the Ly$_\alpha$ frequency;
the expression for $J_\alpha^{C}$ is given in Pritchard and Loeb (2012). When the collisions of neutral hydrogen atoms between themselves and of electrons with protons are taken into account, we have  
\begin{equation}
		x_c=\frac{T_*}{A_{01}T_\gamma}(\kappa^{HH}_{1-0}(T_K)n_H(1-x)+\kappa^{eH}_{1-0}(T_K)n_Hx),
\end{equation}
where $T_*=0.068$~K, $A_{01}=2.85\times 10^{-15}$, the functions $\kappa^{HH}_{1-0}(T_K)$ and $\kappa^{eH}_{1-0}(T_K)$ are also defined in Pritchard and Loeb (2012), and $n_H$ is the hydrogen number density.

The luminosity of the first stars per unit volume is
	\begin{equation}
		\varepsilon_*(z,\nu)=f_*n_H\varepsilon(\nu)\frac{df_{\rm coll}}{dt},
	\end{equation}
where $\varepsilon(\nu)$ is the number of emitted photons per baryon. We take into account the WF effect in a standard way, by summation over the Lyman resonances and integration over the redshifts of the emitted photons whose wavelength decreased, as a result, to the Ly$_\alpha$ line (Barkana and Loeb 2005). In this case, the spectrum is normalized so that the number of photons per hydrogen atom is $N_\alpha=9690$ for Population II stars, while the “recycling fraction” $f_{rec}(n)$ is taken from Pritchard and Furlanetto (2006) and Hirata (2006). Thus, the background of UV photons entering into (\ref{xaleq}) can be written as 
	\begin{equation}
		J_\alpha=\sum\limits_{2}^{30}f_{rec}(n)\!\!\int\limits_z^{z_{\rm max(n)}}\!\! dz'\frac{(1+z)^2}{4\pi}\frac{c}{H(z)}\varepsilon_*(z',\nu'),
	\end{equation}
where $\nu'=\nu(1+z')/)(1+z)$, $(1+z')/(1+z)=[1-(n+1)^{-2}]/(1-n^{-2})$ and $\rm max(n)=30$. On the whole, our approach coincides with that from Mesinger et al. (2011) and Furlanetto (2006) with the correction for the fact that we use the perturbation spectrum with a bump that has the same shape as that in Tkachev et al. (2024a).

Whereas the WF effect determines the boundary of the absorption trough at high $z$z, the gas heating and ionization determines this boundary at low $z$. The heating and ionization are also determined by the first stars and galaxies, but these processes at $z<10$ are very chaotic and have not yet been clarified completely. Therefore, in this paper we focus our attention only on the boundary of the absorption trough at high $z$ and take into account the processes at low $z$ approximately. For example, we specify the contribution to the change in the kinetic gas temperature in the form (Furlanetto 2006)
\[
\frac{2}{3}\,\frac{\varepsilon_X}{k_BnH(z)}=    
\]
\begin{equation}
= 10^3f_X\left(\frac{f_*}{0.1}\,\frac{f_{X,h}}{0.2}\,\frac{df_{coll}/dz}{0.01}\,\frac{1+z}{10}\right)\!\mbox{~K},    
\label{23varep}
\end{equation}
where $k_B$ is the Boltzmann constant and $H(z)$ is the redshift dependence of the Hubble constant. The quantity (\ref{23varep}) enters into the equation for the evolution of the kinetic temperature
	\begin{equation}
		\frac{dT_K}{dt}=\frac{2T_K}{3n_H}\frac{dn_H}{dt}+\frac{2}{3k_B}\sum\limits_j\frac{\varepsilon_j}{n_H}
  \label{dtdteq}
	\end{equation}
together with the contribution of the heating through Compton scattering by CMB photons.    

Large-scale motions of dark matter and baryons accompanied by shocks and gas heating arise during the formation of virialized objects (as they are isolated from the Hubble flow) (Nath and Silk 2001). The shock propagates outside the collapsing object and heats the gas in its vicinity. Although the importance of the gas heating by shocks was pointed out in previous studies in numerical simulations (Gnedin 2004) and analytical calculations (Furlanetto and Loeb 2004), subsequent papers showed that the contribution of shocks is insignificant (see Mesinger et al. (2011) and references therein). Therefore, in this paper, following Mesinger et al. (2011),
we ignore the heating by shocks.

The evolution of the ionization fraction $x$ of a homogeneous gas is calculated from the formula (Furlanetto 2006)
	\begin{equation}
	\frac{dx}{dt}=\zeta\frac{df_{\rm coll}}{dt}-\alpha_rx^2n_H,
	\label{ioneq}
\end{equation}
where $\alpha_r=4.2\times10^{-13}$~cm$^3$~s$^{-1}$ is the recombination coefficient, $\zeta=A_{\rm He}f_*f_{\rm esc}N_{\rm ion}$, $A_{\rm He}$ is the helium fraction, and
$N_{\rm ion}=0.45N_\alpha$ for Population II stars.

	\section{RESULTS OF OUR NUMERICAL CALCULATIONS}
	\label{ressec}
	
Many of the quantities in our calculations are very uncertain. In particular, as shown above, there is the ionization crisis problem. This means that the quantities responsible for the ionization cannot yet be fixed reliably. In this paper we vary $f_{\rm esc}$ and $f_*$ in some range. It can be seen on the upper and lower panels of Fig.~2 that, in addition to the perturbation spectrum, these quantities also affect the result. It can be seen from Fig.~2 that large $f_*$ contradict the observations, since in this case the reionization epoch is shifted to redshifts $z>6.4$ (the transition to zero absorption corresponds to $x=1$ and $\delta T=0$), with this problem being more pronounced for the perturbation spectrum with a bump. However, at small $f_*$ and $f_{\rm esc}$ satisfactory agreement with the observed position of the reionization epoch can be achieved both in the presence and in the absence of a bump.

\begin{figure}[h]
		\centering
		\includegraphics[width=0.47\textwidth]{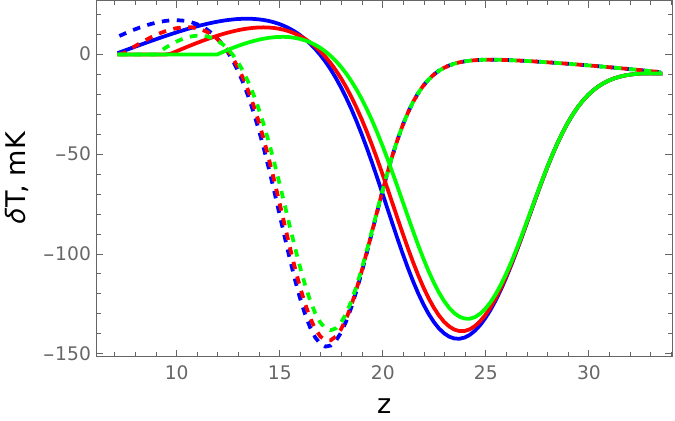}\\
            \includegraphics[width=0.47\textwidth]{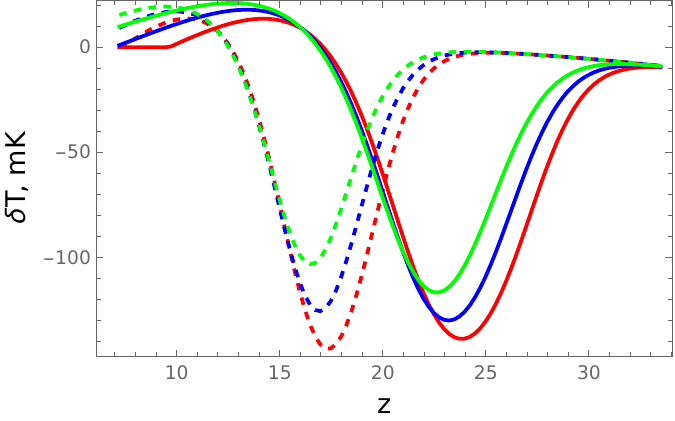}
		\caption{
        The temperature shift due to the absorption in neutral hydrogen. \textit{Top}: in the presence of a bump with $A=20$, $k_0=4.69$~Mpc$^{-1}$ and $\sigma_0=0.1$ (the $gauss\_1$ model) (solid curves) and without a bump (the standard $\Lambda$CDM model) (dashed curves). The parameter $f_{\rm esc}=0.025$, 0.05, and 0.1 (the blue, red, and green curves, respectively). In all three cases $f_*=0.1$. \textit{Bottom}: the same, but for $f_{\rm esc} = 0.05$ and three cases: $f_*=0.025$, 0.05, and 0.1 (the green, blue, and red curves, respectively).
        }
\label{fig2}
\end{figure}
    	
As an example, Fig.~3 shows the result of our calculation in comparison with the EDGES data (Fig.~2 in Bowman et al. (2018)).

\begin{figure}[h]
	\centering
	\includegraphics[width=0.47\textwidth]{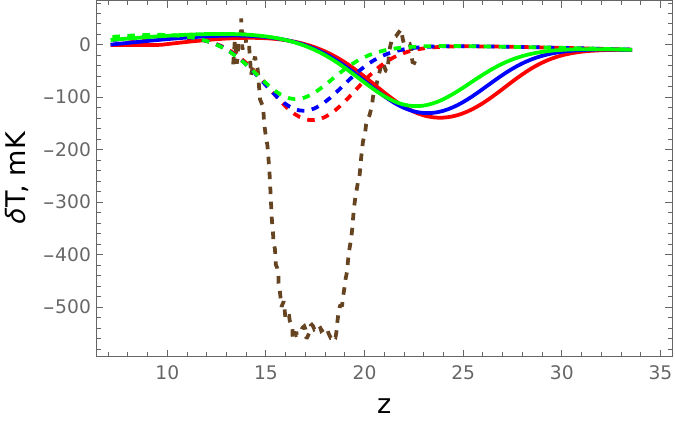}
	\caption{
    Comparison of the EDGES data (the brown dashed curve, Bowman et al. 2018) and our simulations with  $f_{\rm esc}=0.05$ and $f_*=0.025$, 0.05 and 0.1.}
	\label{grdtzvedges}
\end{figure}

The general conclusion of our paper, which may turn out to be useful in future as the physics of the formation of the first stars and the influence of their UV radiation is refined, is the shift in the position of the absorption trough to high redshifts. Instead of $z\sim17$, the position of the minimum in the presence of a bump (with the above parameters) is displaced to $z\sim24$, while the boundary of the absorption trough is shifted to even higher redshifts. This implies that the observation of this feature in the 21~cm line absorption spectrum in the presence of a bump will require radio observations with slightly longer wavelengths and appropriate antenna arrays and noise filtering algorithms, which is a very difficult and nontrivial task in such observations. If meter-band observations will reliably show the presence of an absorption trough with some position of its boundaries and depth, then this will make an independent accurate determination of the perturbation spectrum on the scale of low-mass galaxies in which the first stars were born possible in principle. In this case, refining the various physical processes associated with baryon physics will remain a topical theoretical problem: the formation of the first stars, their properties, and the impact of their UV radiation on neutral hydrogen outside the galaxies.

	\section{DISCUSSION}

In this paper we investigated the CMB absorption in the 21~cm line of neutral hydrogen at the cosmological epoch of Dark Ages in the presence of additional power (bump) in the density perturbation spectrum. The observation of an unexpectedly large number of early galaxies by JWST served as a motivation for modifying the spectrum and introducing an additional bump in Padmanabhan and Loeb (2023) and Tkachev et al. (2024a). The bump in the spectrum leads to the earlier formation of galaxies and, thus, can explain these observations. However, of course, the presence of a bump may also be expected independently of the observations of early galaxies. For example, features in the form of a bump arise in some inflation models. The emergence of primordial black holes is also associated with the possibility of the existence of bumps (for a review of the models, see Inomata et al. (2023) and the recent paper of Lukash and Mikheeva (2025)). Within the known constraints, the shape of the bump and its location so far remain free parameters.
    
A study with similar goals was carried out by Naik et al. (2025), who also analyzed the influence of the bump in the density perturbation spectrum on the 21~cm line profile. However, our approaches differ. First, the amplitude of the bump investigated in this and our earlier papers (for references, see above) is approximately a factor of 10 larger than that in Naik et al. (2025). Second, in this paper, to construct the mass function of gravitationally bound halos (galaxies), we performed numerical simulations rather than approximate analytical calculations (the Press–Schechter formalism (Press and Schechter 1974) and its elaboration in subsequent papers (Sheth and Tormen 2002)). The $gauss\_1$ model as an extension of the standard cosmological model has a specific status—it describes the abundance of massive galaxies at high $z$ better than does the standard model.

The absorption in the 21~cm line at the epoch of Dark Ages is sensitive to many processes. Significant uncertainties related to the formation of the first stars and the efficiency of the impact of their UV radiation on the medium remain even within the standard baryon physics. For example, among the more exotic processes, the decay or annihilation of dark matter particles could affect the spin temperature and the shape of the 21~cm absorption trough (Novosyadlyj et al. 2024). In this paper we assume standard microphysics at the epoch of Dark Ages and vary only the perturbation spectrum by assuming the presence of an additional bump in the form of a narrow Gaussian, as suggested in Tkachev et al. (2024a). Apart from two models with bumps, which yielded virtually identical results with regard to the 21~cm line, we also considered the case where a distributed excess is present in the density perturbation spectrum on small scales (the so-called blue spectra, namely the $b-tilt\_k10$ model from Tkachev et al. (2024b)). In this case, however, the stars are formed early, and their radiation strongly ionizes and heats the gas; as a result, there is no 21~cm line absorption in the $b-tilt\_k10$ model. The microphysics at the perturbation formation epoch, i.e., the shape of the inflation potential that is not yet fixed unambiguously in the theories of elementary particles, is responsible for the appearance of features in the form of broad or narrow bumps in the smooth density perturbation spectrum.

In contrast to Naik et al. (2025), to compute the formation of the first stars, we use not the linear theory of evolution of perturbations built in the \texttt{21cmFAST} software package, but apply the results of full scale numerical simulations of the formation of early structures. As pointed out above, we also use a significantly different bump model with the parameters needed to explain the JWST observations. As a criterion for the selection of dark matter halos in which the first stars are formed, we use the criterion for the onset of gas cooling at a virial temperature $>10^4$~К. Nevertheless, our computations qualitatively confirm the conclusion of Naik et al. (2025) that the absorption trough in the 21~cm line is shifted
toward high redshifts due to the earlier formation of structures and the first stars giving UV radiation, although a direct comparison with the results of Naik et al. (2025) is impossible due to the use of a different bump model.

In future, when the mechanisms of the formation of the first stars and the reionization of the Universe will be clarified in more detail, the absorption in the 21~cm line can become an accurate tool for determining the initial perturbation spectrum in the range of masses corresponding to galaxies. It will be possible to observe the global absorption signal both with bolometric telescopes analogous to EDGES and SARAS (Singh et al. 2022), for example, using the telescope being planned (de Lera Acedo 2019), and, possibly, with giant radio telescopes like Square Kilometre Array (SKA) being put into operation or being planned, especially at their long-wavelength subsystems. At this juncture, however, this method can give only some qualitative and preliminary conclusions due to the existence of large theoretical uncertainties and problems. The uncertainties concern, in particular, the formation efficiency of the first stars. The uncertainty in the fraction of baryons collapsing into stars, $f_*$, introduces a significant uncertainty into the calculations of the absorption in the 21~cm line. This paper does not purport to make accurate quantitative predictions, but demonstrates the influence of the bump in the perturbation spectrum on the result.
	
The problem that takes place even without an additional bump is the overproduction of ionizing photons (Munoz et al. 2024). In the presence of a bump and at constant other parameters describing the baryon physics, the reionization epoch is shifted to even higher redshifts compared to the models without a bump. Without an additional modification of the model, this comes into conflict with observations. Just as in the case without a bump (but more radically), we must assume that the hydrogen ionization efficiency was suppressed through an as yet unknown effect. Further observations in this direction are required to clarify this situation. As
a problem we may also point to the inconsistency of the available EDGES observational data on the absorption in the 21~cm line with the calculations of the absorption trough depth. The EDGES data show a significantly larger depth than that given by the theoretical calculations (both without and with a bump), while the position of the absorption trough on the redshift scale corresponds better to the model without a bump. However, it should be noted that the EDGES observational data on strong absorption were not confirmed by the observations of another radio telescope, SARAS 3, in which the absorption trough was invisible (Singh et al. 2022). At the same time, because of the large observational errors, the SARAS 3 data are consistent with the results of our calculations of the absorption trough depth and position, both in the presence and the absence of a bump. Therefore, the depth and position of the absorption trough are still far from being clarified from an observational point of view, and new, more accurate studies are also required here.

    \section{ CONCLUSIONS}

The radiation of the first stars reduces the spin temperature of neutral hydrogen, amplifying the absorption in the 21~cm line. A comparison of the cosmological models with and without a bump in the density perturbation spectrum shows that it is possible to determine the probable position of the bump in the perturbation spectrum and, thus, to reconstruct the spectrum of cosmological perturbations on scales $k>1$~Mpc${}^{-1}$ from the position of the absorption frequency profile. For the analysis we used the results of our numerical N-body simulations. The parameters of the bump were chosen so as to correspond to the excess of galaxies at $z\sim10$ observed by JWST.    

We found a significant shift of the absorption trough boundary in the presence of a bump (Fig.~2), which is a sensitive tool in reconstructing the power spectrum on small scales. Our calculations agree qualitatively with the conclusion of Naik et al.~(2025) about the position of the absorption trough at high redshifts.

A number of new radioastronomical observations in the meter band with various radio telescopes are expected in the near future. If these observations will allow one to adequately reveal the absorption trough and to describe its boundary, then this will give valuable information about the shape of the cosmological density perturbation spectrum in the early Universe. The absorption in the 21~cm hydrogen line at $z>10$ is an accurate tool for the construction of a cosmological model on small scales.
	    
    \section*{ACKNOWLEDGMENTS}

We are grateful to the referee for the useful remarks that improved the content of the paper.

    \section*{FUNDING}
    
The work of V.N. Lukash, E.V. Mikheeva, S.V. Pilipenko, and M.V. Tkachev was performed within the State assignment of the Astro Space Center of the P.N. Lebedev Physical Institute. The work of Yu.N. Eroshenko was performed within the State assignment on the theme of the Institute for Nuclear Research.

    \section*{CONFLICT OF INTEREST}

The authors of this work declare that they have no
conflicts of interest.
	
    
\end{document}